\documentstyle[11pt,newpasp,twoside,psfig]{article}
\begin{document}
\title{The INT-WFS + 2dFGRS: The Local Space \& Luminosity Density.}
\author{Nicholas Cross, Simon Driver \& David Lemon}
\affil{School of Physics and Astronomy, North Haugh, St Andrews, Fife, 
KY16 9SS, United Kingdom.}

\begin{abstract}

We discuss the quantification of the local galaxy population and the impact 
of the ``New Era of Wide-Field Astronomy'' on this field, and, in 
particular, systematic errors in the measurement of the Luminosity Function.
New results from the 2dFGRS are shown in which some of these selection effects
have been removed. We introduce an INT-WFS project which will further reduce 
the selection biases. We show that there is a correlation between the surface 
brightness and the luminosity of galaxies and that new technologies are having
a big impact on this field. Finally selection criteria from different surveys 
are modelled and it is shown that some of the major selection effects are 
surface brightness selection effects.
   
\end{abstract}
\vspace{-5mm}
\section{Introduction}

Galaxy populations were first studied by Hubble (1926), who developed the 
familiar Tuning Fork diagram of Ellipticals, Spirals and Barred Spirals. Most 
bright galaxies can be morphologically classified by their Hubble type. 
However, many types of galaxy have been found that don't fit the Tuning Fork. 
These occur both at low redshift and at high redshift where the galaxies can 
be intrinsically different due to evolution. Some of these galaxies are shown 
in Fig. 1. The Tuning Fork can be extended to include these new types of 
galaxy as shown in Fig. 2. 

\begin{figure}[h]
\psfig{file=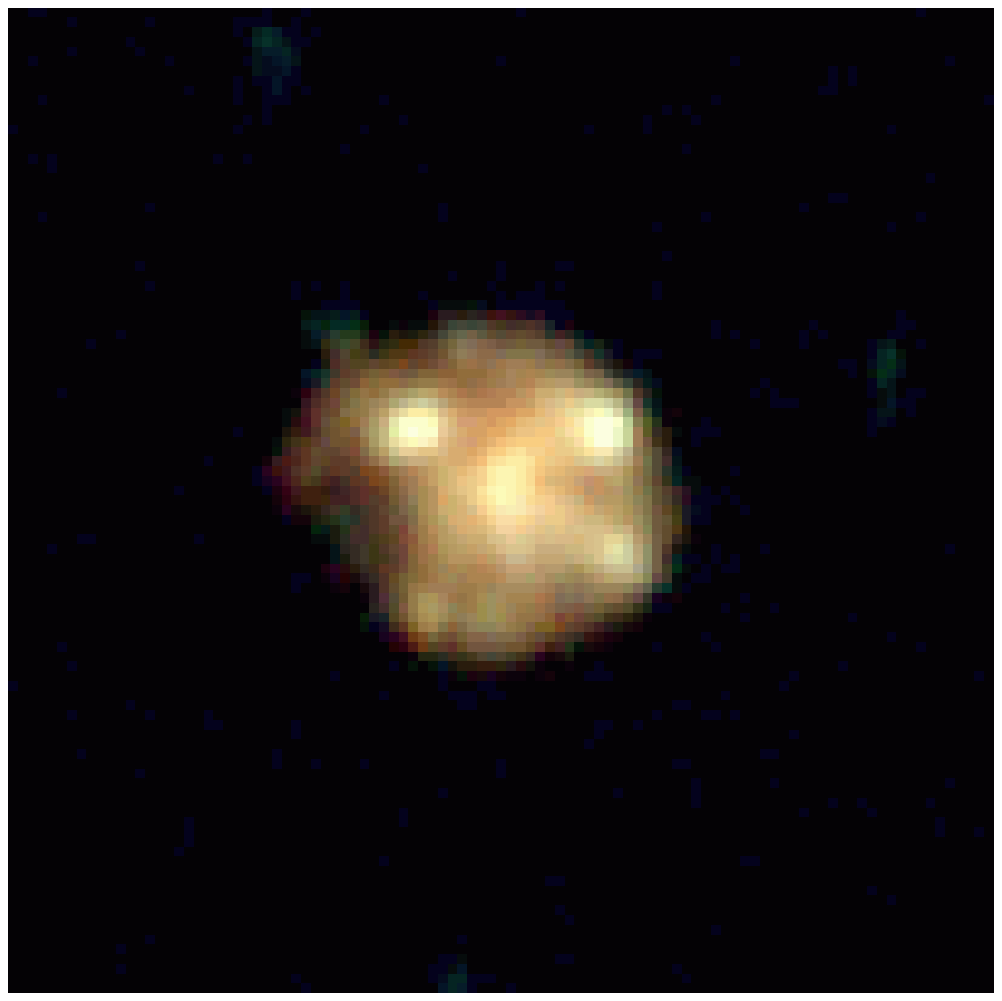,height=25.0mm,width=25.0mm}
\vspace{-25.0mm}
\hspace{30.0mm}
\psfig{file=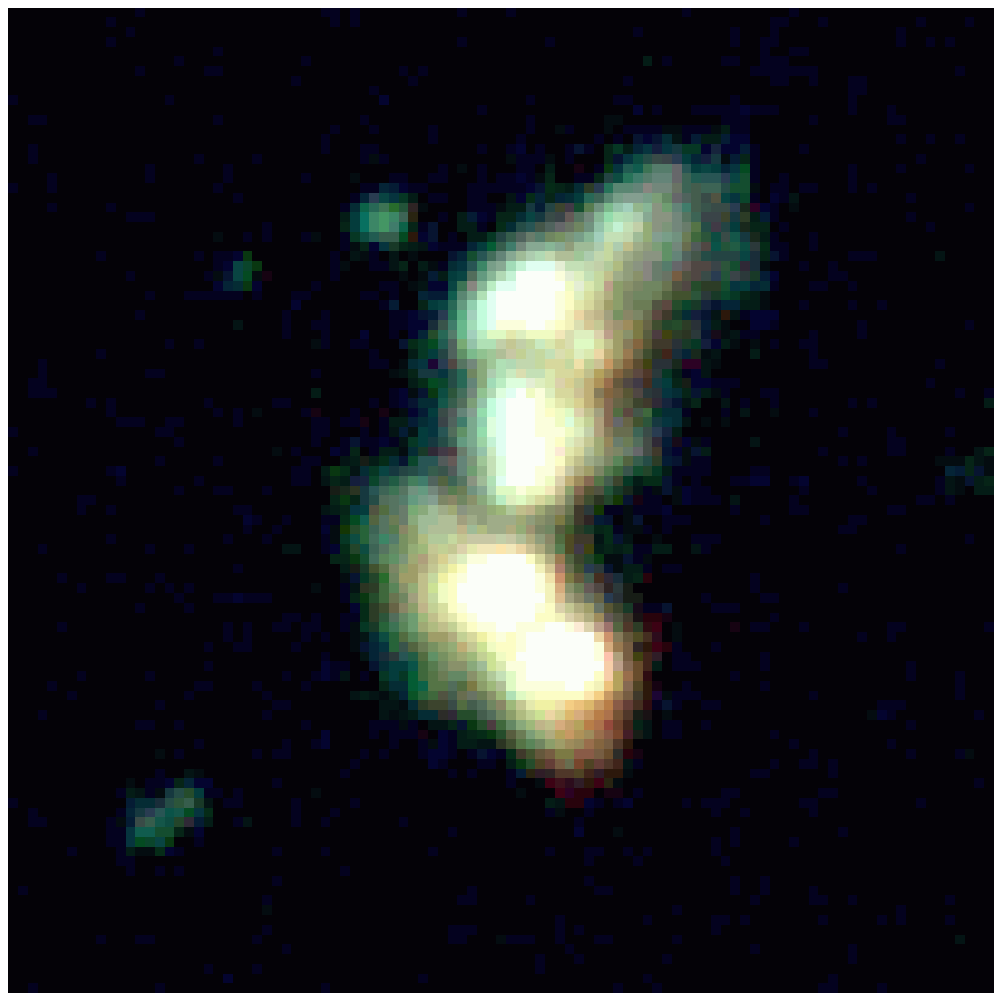,height=25.0mm,width=25.0mm}
\vspace{-25.0mm}
\hspace{5.0mm}
\psfig{file=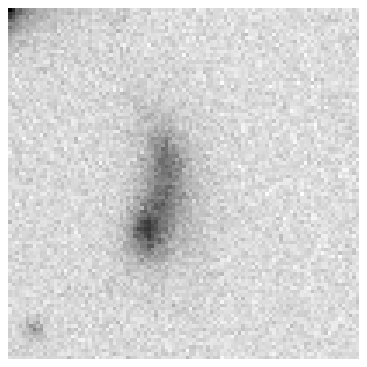,height=25.0mm,width=25.0mm}
\vspace{-25.0mm}
\hspace{5.0mm}
\psfig{file=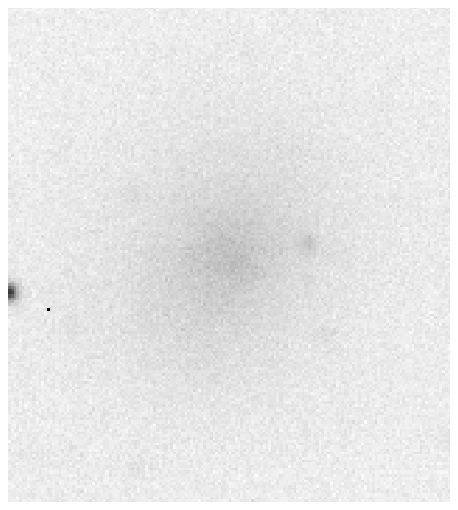,height=25.0mm,width=25.0mm}
\vspace{50.0mm}
\caption{Some peculiar galaxies. The two on the left hand side are high 
redshift galaxies in the Hubble Deep Field North. The two on the right are low
redshift galaxies from the Millennium Galaxy Catalogue. The farthest right
galaxy is a Low Surface Brightness Galaxy.}
\end{figure}

\begin{figure}
\centerline{\psfig{file=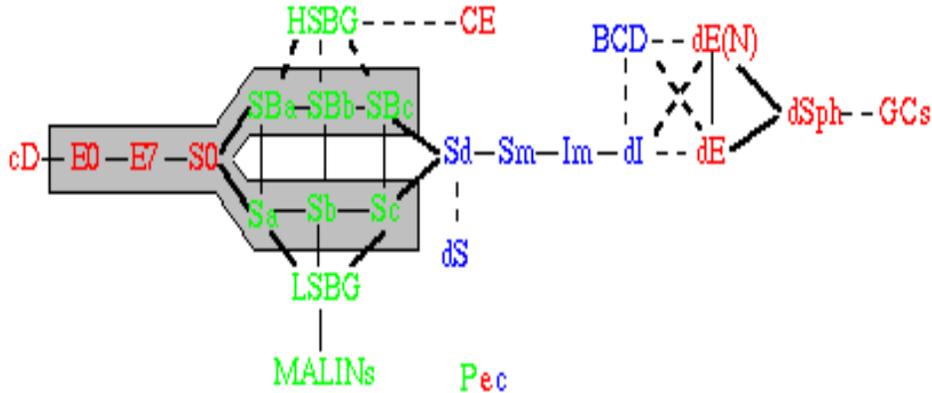,height=55.0mm,width=130.0mm}}
\caption{The various different classes of galaxies that astronomers have named
over the years. The classes are positioned roughly in order of magnitude 
from bright on the left to faint on the right and in surface brightness from 
high surface brightness at the top to low surface brightness at the bottom. The
Hubble Tuning Fork types make up most of the bright galaxies.}
\vspace{-5mm}
\end{figure}

However, morphological classification only tells us what type of galaxies there
are. It does not tell us what proportion are of each type or whether this 
varies over time. Galaxy formation and evolution is a relatively new field and
so far very little is known about what produced the galaxy population we see
today. To understand this process it is first necessary to have
some quantitative information about the local galaxy population. The 
luminosity is an easy to measure quantity which can be used to classify 
galaxies. The Luminosity Function (LF) (Peebles \& Hauser 1974) 
measures the space density of galaxies as a function of luminosity. This can 
be convolved with different evolutionary models and compared to number - 
magnitude counts (Driver et al. 1994). However, recent surveys have produced a
large range in the measured LF, Fig. 3, with the variation between the surveys
much greater than random errors. The systematic errors causing this wide 
variation must be understood before any progress is going to be made.

\begin{figure}
\centerline{\psfig{file=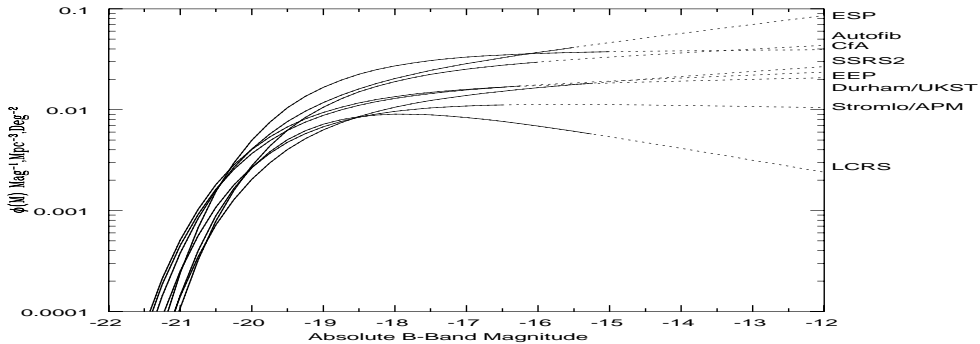,height=55.0mm,width=130.0mm}}
\caption{This is a plot of eight recent Galaxy Luminosity Functions. The
variation at the M$^*$ ($M\approx-19.5$) point is a factor of 2 and at 
$M=-14.5$ the variation is a factor of 10. The $M^*$ point is where the 
statistics are best. The variation would be expected to be of the order 10\%
if it where due to random errors. For more details see Cross et al. (2000).}
\vspace{-5mm}
\end{figure}

Disney (1976) pointed out that surface brightness selection effects are 
important to take into account when measuring the luminosity function. 
Phillipps \& Disney (1986) and Phillipps, Davies \& Disney (1990) went on to 
calculate the effects of surface brightness. These calculations take into 
account light lost below the isophote, Malmquist bias and surface brightness 
dimming due to cosmological expansion. 

Here, we will discuss how we have used the ``Two Degree Field Galaxy Redshift 
Survey'' (2dFGRS) to produce a Bivariate Brightness Distribution (BBD), which 
corrects for these effects. Then we will describe some of the results and some
of the selection effects that we have not corrected for. We will then briefly 
describe the Millennium Galaxy Catalogue (MGC), one of the projects in the 
Isaac Newton Telescope Wide Field Survey (INT-WFS). The MGC is a deep, imaging
survey designed to remove some of the remaining selection effects. The 
selection effects are then modelled by convolving the BBD with different 
functions based on the visibility theory of Phillipps, Davies \& Disney 1990) 
adopting parameters appropriate for the ESO Slice Project (ESP) (Zucca et al. 
1997), the Las Campanas Redshift Survey (LCRS) (Lin et al. 1996) and the Second
Southern Sky Redshift Survey (SSRS2) (Marzke et al. 1998). 

\section{The 2dFGRS}

The 2dFGRS is a $2000\Box^{\circ}$ redshift survey of the North Galactic and 
South Galactic Polar (NGP, SGP) regions. When it is complete it will have 
redshifts for 250,000 galaxies with $B<19.45$. At present more than 
100,000 redshifts have been measured, making this the largest survey to date 
by a factor of 4. The input catalogue is the APM catalogue (Maddox et al. 
1990a,b). For more details on the 2dFGRS see the article in these proceedings
by Bridges et al.

\subsection{The Bivariate Brightness Distribution (BBD).}

One method to remove surface brightness selection effects is to construct a 
BBD. This is a plot of the space density of galaxies versus the absolute 
magnitude, AND the effective surface brightness. This is a simple extension 
of the LF, whereby galaxies of different surface brightnesses are treated 
separately. In Cross et al. (2000),
we describe how to construct the BBD for the 2dFGRS. Throughout this work, we 
make corrections based upon the data as much as possible. The data has been 
corrected for light lost below the isophote, Malmquist bias, redshift 
incompleteness and clustering. In each case these corrections were functions 
of both absolute magnitude AND effective surface brightness. The BBD was 
produced from 50,000 galaxies in the SGP.

\subsection{Results} 

Fig. 4 shows the final Malmquist corrected space density of galaxies as a 
function of $M_B$ and $\mu_e$. There is a strong luminosity-surface brightness
relation. L-$\Sigma$ relations
are also seen in Driver (1999), Binggeli (1993) and de Jong and Lacey (2000). 
The shaded region on Fig. 4b shows where the volume is less than 
10,000Mpc$^3$. This plot suggests that luminous LSBGs such as Malin 1 are
extremely rare. Faint HSBGs such as M32 are likely to be rare as they are far
from the L-$\Sigma$ axis.

\begin{figure}[h]
\psfig{file=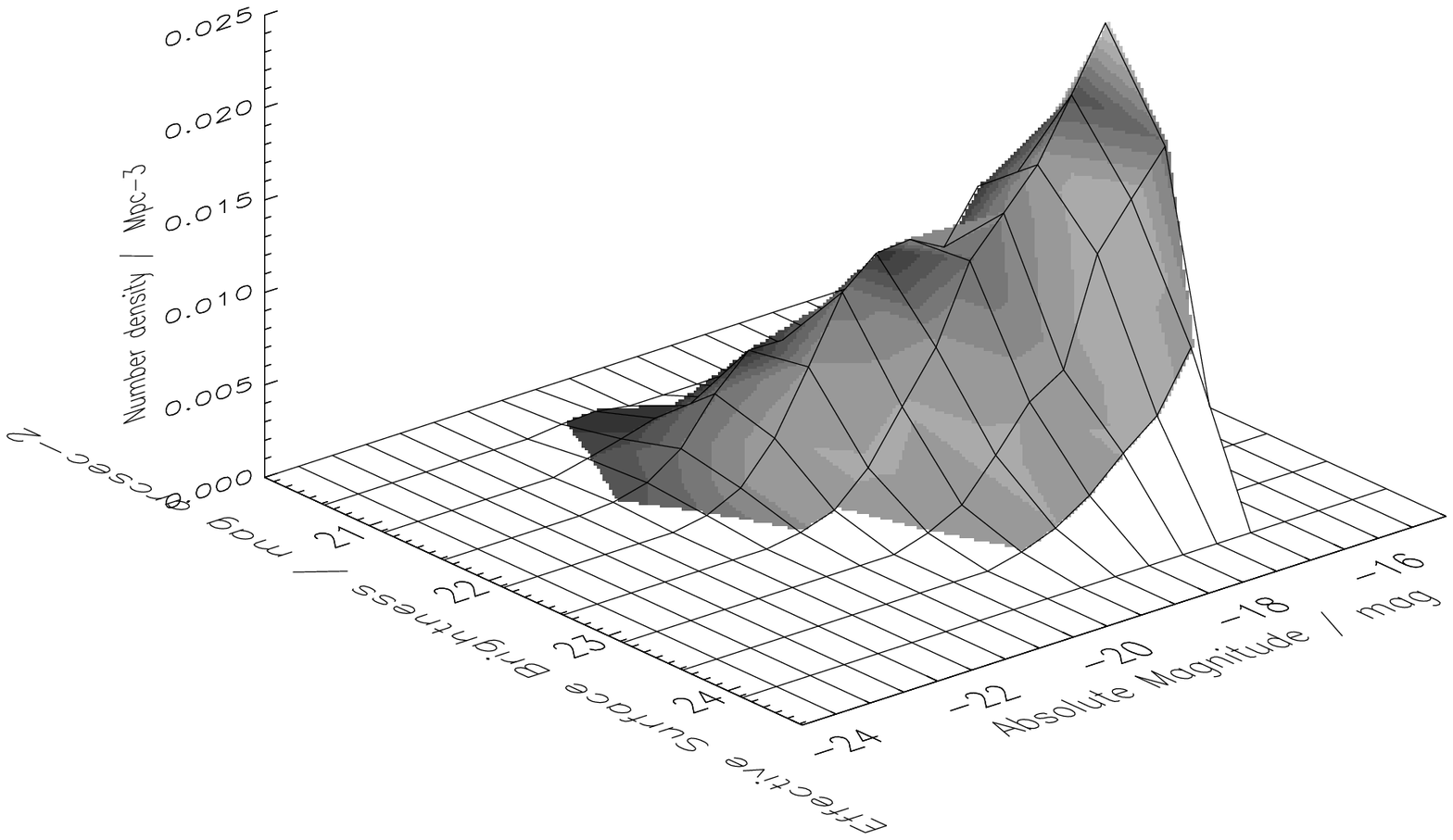,height=55.0mm,width=55.0mm}
\vspace{-55.0mm}
\hspace{65.0mm}
\psfig{file=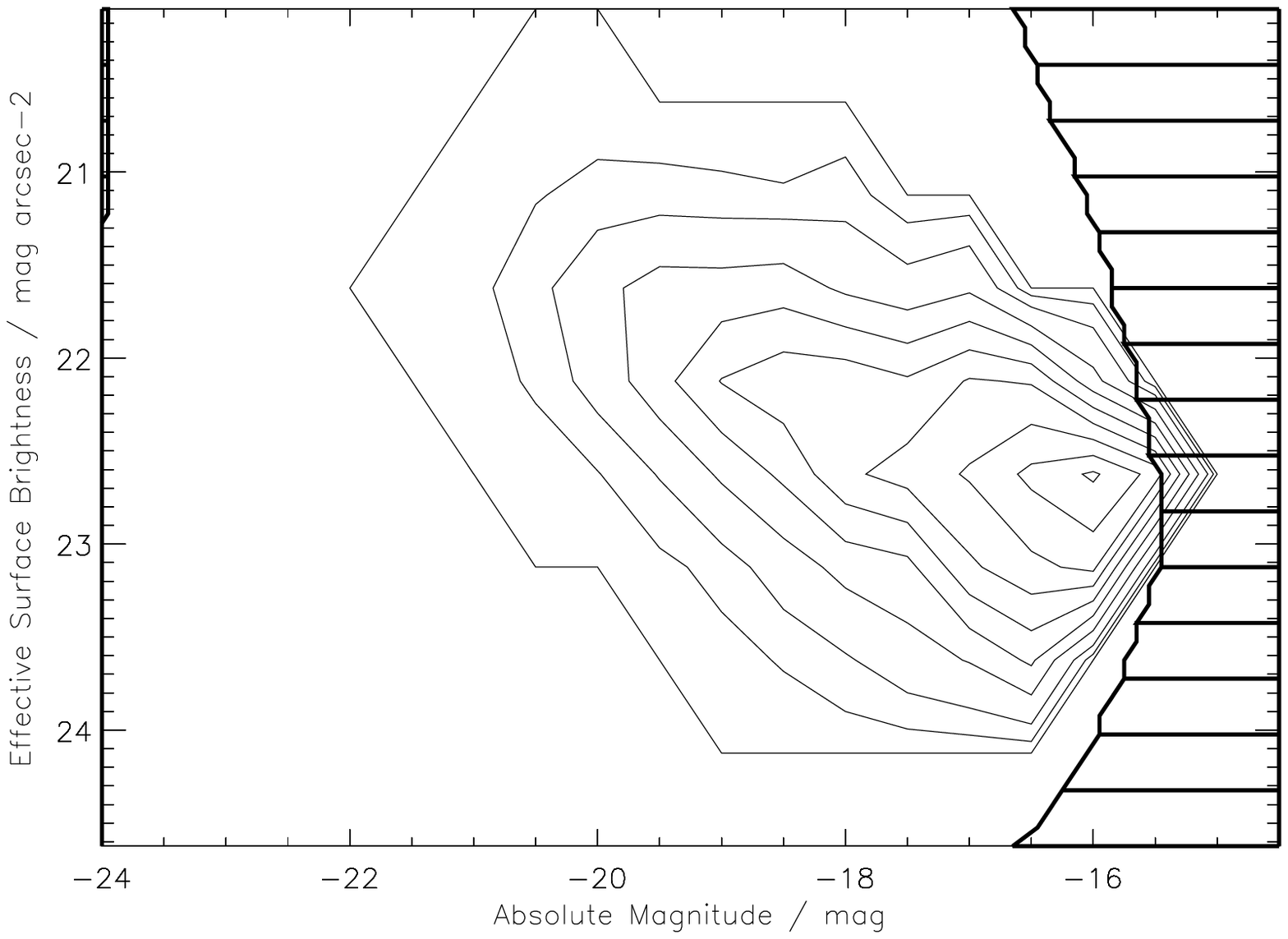,height=55.0mm,width=55.0mm}
\caption{The 2dFGRS space density. These plots 
are reproduced from Cross et al. (2000). The contour levels are 
$1.0\times10^{-7}$, $1.0\times10^{-3}$, $2.5\times10^{-3}$, $5.0\times10^{-3}$,
$7.5\times10^{-3}$, $1.0\times10^{-2}$, $1.25\times10^{-2}$, 
$1.5\times10^{-2}$, $1.75\times10^{-2}$, $2.0\times10^{-2}$, and
$2.25\times10^{-2}$ galaxies Mpc$^{-3}$bin$^{-1}$ in the right hand plot.} 
\vspace{-5mm}
\end{figure}

\begin{figure}[h]
\psfig{file=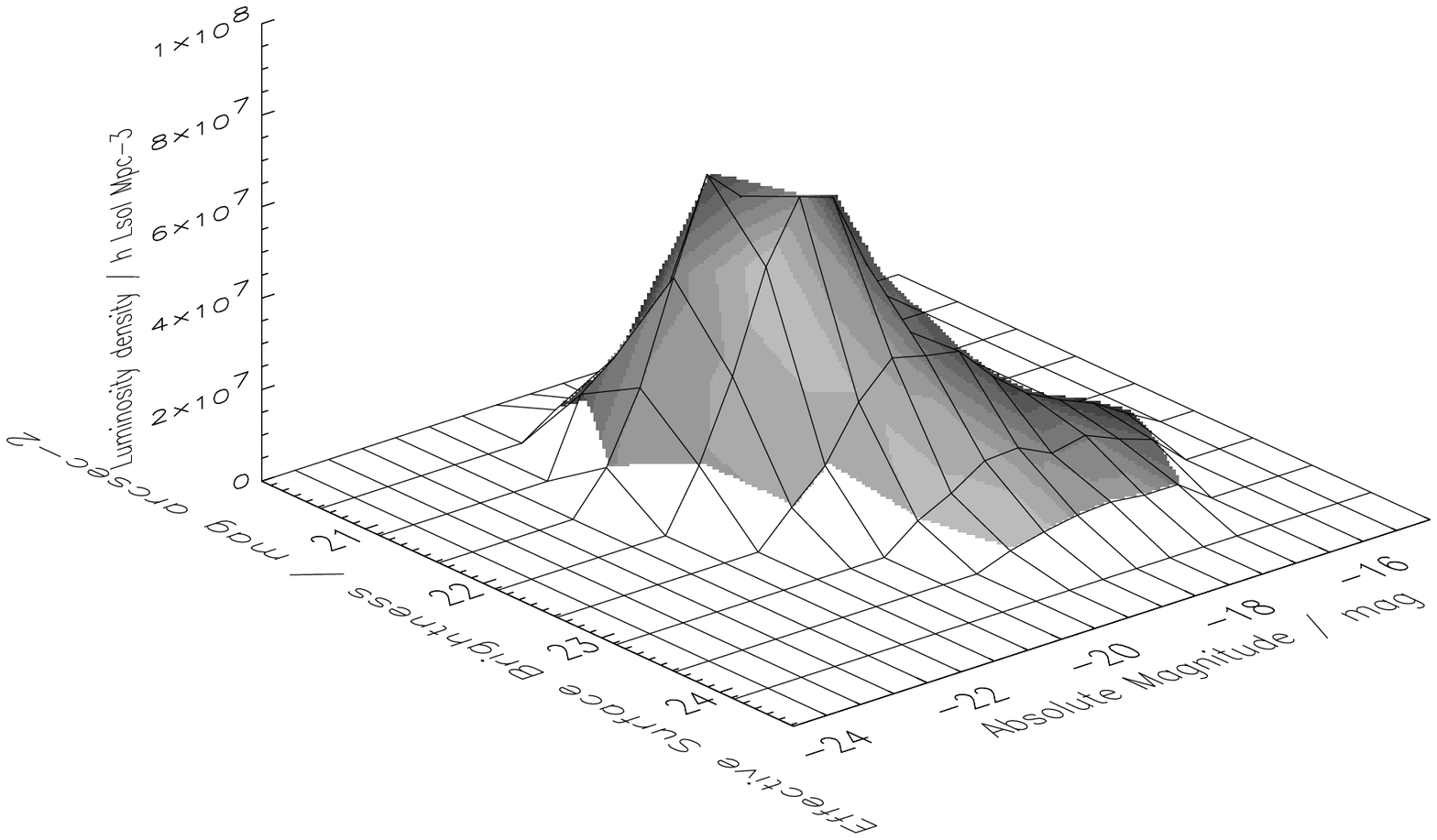,height=55.0mm,width=55.0mm}
\vspace{-55.0mm}
\hspace{65.0mm}
\psfig{file=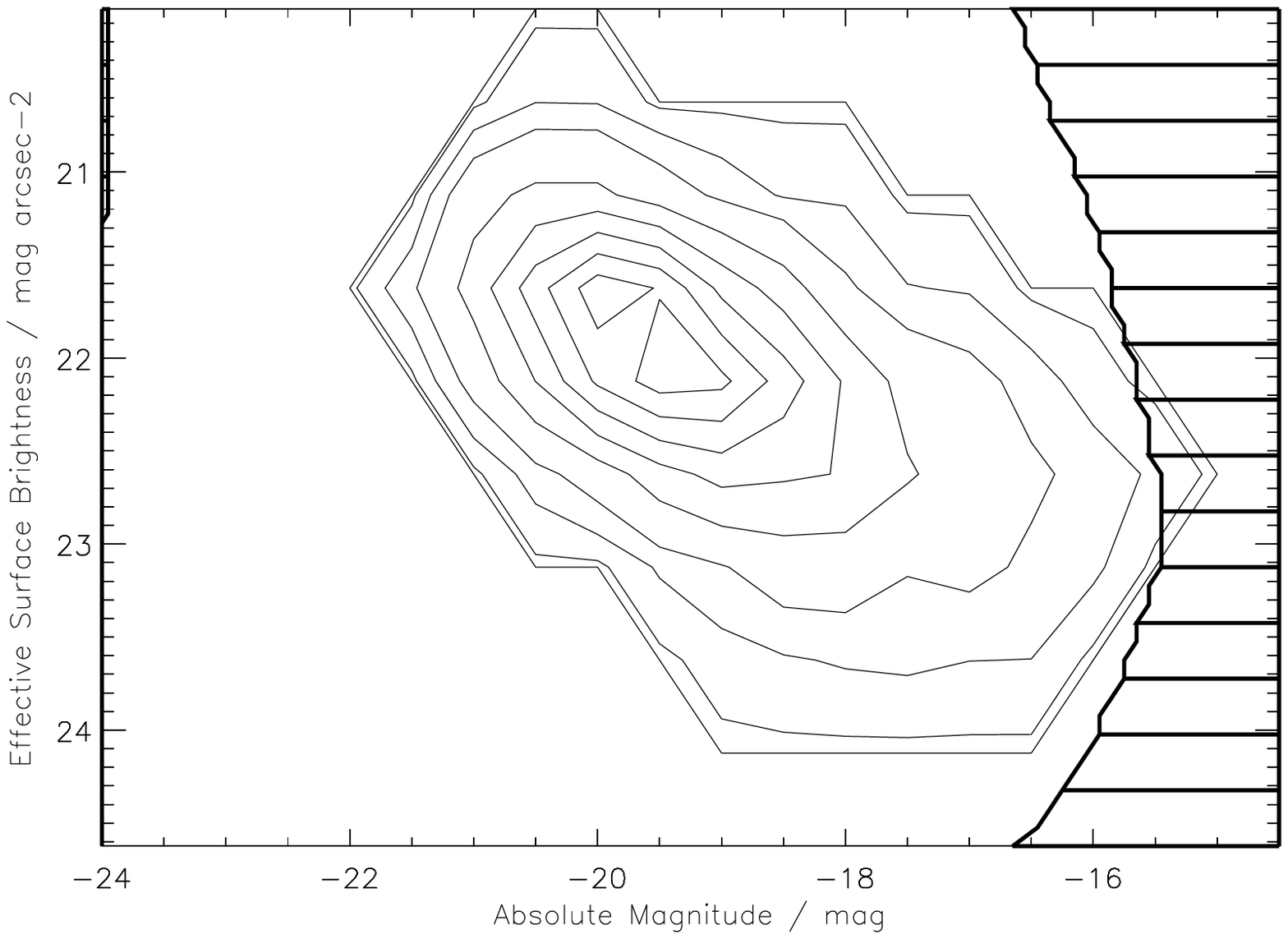,height=55.0mm,width=55.0mm}
\caption{The 2dFGRS luminosity density. These  plots 
are reproduced from Cross et al. (2000).
The contour levels are $100$, $1.0\times10^{6}$, $5.0\times10^{6}$, 
$1.0\times10^{7}$, $2.0\times10^{7}$, $3.0\times10^{7}$, $4.0\times10^{7}$, 
$5.0\times10^{7}$, and $6.0\times10^{-2}$L$_{\odot}$Mpc$^{-3}$bin$^{-1}$ in 
the right hand plot.}
\vspace{-5mm}
\end{figure}

The luminosity density $j=2.62\pm0.20\times10^8h_{100}L_{\odot}$Mpc$^{-3}$.
The luminosity density is plotted as a function of absolute magnitude and 
surface brightness in Fig. 5. This shows clearly that the luminosity density
is decreasing well before the selection boundaries imposed by the survey. 
This implies that we are indeed seeing the majority of the light emitted by 
local galaxies. 

\subsection{Problems with the 2dFGRS BBD.} 

The input catalogue completeness is a complex function of both surface 
brightness and magnitude. The curve on Fig. 4b shows where the volume, over 
which a galaxy can be detected, is equal to 10,000Mpc$^3$. Outside this 
region, the detection of galaxies becomes increasingly difficult. The input 
catalogue is only complete in regions determined by the detection isophote 
($24.7$mag arcsec$^{-2}$), the apparent magnitude limit ($19.45$mag) and the 
smallest isophotal radius ($3.6''$). The smallest isophotal radius depends
on the seeing ($>2''$ for the APM). 

The APM suffers from less than ideal photometry which is only accurate to 
$\sim0.2$mag. The data does not contain information on the light profiles of 
the galaxies such as disc-bulge separation parameters.

\section{The Millennium Galaxy Catalogue.}

To remove these problems we have undertaken a deep equatorial CCD survey over 
part of the 2dFGRS Northern Galactic Polar region. This survey is described
in more detail in these proceedings by Lemon et al. It will push the 
boundaries of the $M-\mu_e$ plane that can be studied, giving us a 
quantitative measure of the population that we may have missed with the 
2dFGRS. It will also improve the 
photometry of the galaxies we have already observed and allow us to do 
morphological classification and disc-bulge separation.
\vspace{-5mm}
\section{Modelling Selection Effects}

Fig. 6 shows the Schechter function fits to three LFs shown 
in Fig. 3. These are the LCRS (Lin et al. 1996), the SSRS2 (Marzke et al. 1998)
and the ESP (Zucca et al. 1997). The LCRS and ESP show the largest variation 
amongst the surveys. Also plotted are three lines showing the LFs produced 
when the 2dFGRS BBD is convolved with three different visibility functions. 
The parameters used for the LCRS are $\mu_{lim}=23.3$B$\mu$, m$_{lim}=19.2$B,
and d$_{min}=4.0$, calculated using a B-r colour of 1.5 (Driver et al. 1994).
The parameters used for the ESP are $\mu_{lim}=25.0$B$\mu$, m$_{lim}=19.4$B,
and d$_{min}=4.0$. The parameters used for the SSRS2 are 
$\mu_{lim}=25.0$B$\mu$, m$_{lim}=15.5$B, and d$_{min}=4.0$. This plot shows 
that the LFs can be reproduced well at the faint end. This demonstrates that 
surface brightness selection effects are a major systematic error in 
determining luminosity functions.

\begin{figure}
\psfig{file=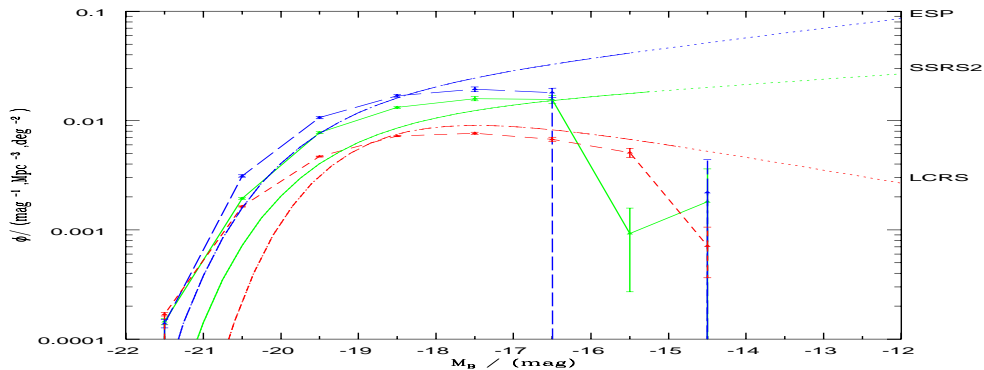,height=55.0mm,width=130.0mm}
\caption{Plot of LFs of recent surveys. The smooth curves are the Schechter 
Function fits to the original surveys. The connected triangular points with 
error-bars are the luminosity distributions calculated by convolving the BBD
with various selection functions. The SSRS2 is the solid line, the LCRS is the
short-dashed line and the ESP is the long-dashed line. The dots represent
where the data runs out for each survey.}
\vspace{-5mm}
\end{figure}

\section{Conclusions}

To understand the effects of evolution, it is essential to have a good 
quantitative knowledge of the local galaxy population.

It is apparent from Fig. 4 that our knowledge of the local galaxy population
is limited by systematic errors rather than random errors. 

A major systematic error is caused by surface brightness selection 
effects. These can be removed using a Bivariate Brightness Distribution.

The new technologies in this ``New Era of Wide Field Astronomy'' allow us to 
detect larger numbers of galaxies with higher precision. These new 
technologies are the multi-fibre spectrometers such as 2dF and the large format
CCDs such as those in the INT Wide Field Camera. These new surveys allow us 
to properly tackle the selection effects that have dogged our surveys and 
therefore will allow us to finally pin down the space and luminosity densities
of galaxies.

\end{document}